\documentstyle[epsfig]{elsart}
\newcommand{\np}[1]{Nucl. Phys. {\bf #1}}
\newcommand{\pl}[1]{Phys. Lett. {\bf #1}}
\newcommand{\pr}[1]{Phys. Rev. {\bf #1}}
\newcommand{\prl}[1]{Phys. Rev. Lett. {\bf #1}}
\newcommand{\zp}[1]{Z. Phys. {\bf #1}}

\begin{document}

\makeatother

\unitlength1cm
\textheight 233mm

\begin{frontmatter}
\title {{\bf Nature of neutrinos in the light of present and future experiments}}

\author{ M. Czakon$^1$},
\author{ J. Gluza$^{1,2}$},
\author{M. Zra\l ek$^1$}

\address{$^1$ Department of Field Theory and Particle Physics, Institute 
of Physics, University of
Silesia, Uniwersytecka 4, PL-40-007 Katowice, Poland}
\address{$^2$ DESY Zeuthen, Platanenallee 6, 15738 Zeuthen, Germany}

\begin{abstract}
Two general models with hierarchical and almost degenerate neutrino masses,
which are able to explain 
the solar and atmospheric anomalies are investigated.  
We show how   neutrinoless double beta decay experiments discern 
Dirac and Majorana 
natures of neutrinos. The strongest result is, that almost degenerate neutrinos 
with masses above 0.22 eV combined with the SMA MSW must be Dirac particles.
In the general case, the same bound is true for specific parity assignments.
In the near future (GENIUS experiment), 
one of the hierarchical schemes, and the VO mechanism for almost
degenerate neutrinos and specific mixing matrix elements and parity assignments
will be tested.  
\end{abstract}
\end{frontmatter}
\section{Introduction}

There are three (Weyl, Dirac and Majorana) types of spin-half fermion fields.
Massless fermions are Weyl particles. Massive spin-half objects can be of either
Dirac or Majorana type. All charged fermions are Dirac particles as a
consequence of the electric charge conservation. Conservation of the lepton number
is decidedly less fundamental than the electric charge conservation.
Without lepton number conservation, neutrinos do not hold any additive internal
`charge' and can be identical to their own antiparticles. Such fermions are
known as Majorana particles. The problem whether neutrinos are of Dirac or
Majorana type has a long history (for a review, see e.g. \cite{zr}). 
The main problem is that
observable effects which could differentiate between them 
disappear continuously with the vanishing of masses for single-handed states.
This is known as the ``Practical Dirac-Majorana confusion
theorem'' \cite{li}.

From the theoretical point of view it is quite likely that neutrinos are
Majorana particles. Such objects are more fundamental and almost all
extensions of the Standard Model predict their existence. Dirac
particles can be considered as composed of two Majorana particles with
opposite CP parities. But even if theory suggests that
neutrinos are presumably identical to their own antiparticles,  such a property should be
checked experimentally.

It might be, that we reached a point where the correct answer is close.

An obvious place is
the neutrinoless double $\beta $ decay, ($\beta \beta $)$_{o \nu}$, of nuclei.
The Heidelberg-Moscow germanium
experiment gives a lower limit on the half-life time \cite{bb}

\begin{equation}
T^{0 \nu}_{1/2} > 5.7 \cdot 10^{25} \mbox{\rm yr (at } 90 \% \; \mbox{\rm C.L.).}
\end{equation}

This result excludes an  effective Majorana neutrino mass greater than 0.2
eV
\begin{equation}
|<m_{\nu}>| < 0.2\; \mbox{\rm eV.}
\end{equation}

The Heidelberg-Moscow collaboration
proposed a new project (GENIUS) which is anticipated to be sensitive on
\cite{kla}: 
\begin{equation}
|<m_{\nu}>| \simeq  0.01\; \mbox{\rm eV}.
\end{equation}

These bounds alone are not enough to deduce the nature of neutrinos. 
They give only restrictions on the combination of Majorana neutrino 
mixing matrix elements $U_{ei}^2$ and their masses $m_i$. 
However, this information can be linked to other  experiments. 

The connection between mass and oscillation properties of neutrinos,
which follows from different experiments, has been discussed in many
papers \cite{previous,b1,raj,geo,ten,eleven}. Howerever, most authors
assume that neutrinos are Majorana particles. In this case, bounds on
the $(\beta\beta)_{0\nu}$ decay give either informations on the solar
neutrino mechanism, or on the neutrino mass scale (see
e.g. \cite{b1,raj,geo}). Inversely, a guess on the solar neutrino
mechanism, together with a neutrino mass scale, yields $\langle m_\nu
\rangle$ (see e.g. \cite{ten}). 

Here we propose a different line of thought. The information on
neutrino matrix elements and masses, given by oscillation experiments
and tritium $\beta$ decay, is independent of the neutrino nature. We
collect this data from the existing experiments and many
phenomenological works \cite{review}, and use it to check whether
bounds from negative $(\beta\beta)_{0\nu}$ results are satisfied. If
they are, neutrinos may be (but not necessarily are) Majorana
particles, if not they must be Dirac particles. In this work,
we discuss present (Eq. (2)), and future bounds (Eq. (3)) on the
$(\beta\beta)_{0\nu}$ decay.

In the next Section we collect all the relevant information on
mixing matrix elements and masses extracted from  experiments
and establish two general mass schemes which are consistent with both solar and 
atmospheric data.
We assume that only three massive neutrinos exist with two different and
independent $\delta $m$^2$. This means that we do not consider the still unsettled
LSND anomaly \cite{lsnd} waiting for its future confirmation  \cite{con}.
The main discussion is given in Section 3, where  we try to infer
the neutrino's character  in the frame of the proposed neutrino schemes. 
Conclusions are to be found in Section 4.

\section{Experimental information on neutrino masses and
mixing matrix elements}

Since we want to discuss
both Dirac and Majorana neutrinos we must introduce appropriate mixing matrices.
Without loss of
generality we work in a basis where the charged lepton mass matrix is
diagonal \cite{aq}. The matrix that relates the flavour Majorana states to the mass
eigenstates may be parameterized by three Euler angles $\theta_i$, i=1,2,3 and
three phases $\delta,  \phi_2$ and $\phi_3$ \cite{b1}. However, only one of the 
three phases  ($\delta $)
has physical meaning to neutrino oscillations, independently of the neutrino
character \cite{aq}. The two other phases ($\phi_2,\phi_3$) enter the 
 ($\beta \beta $)$_{0 \nu}$ amplitude. So, let
us write the mixing matrix in a compact form 
(c's and s's are shortcuts for appropriate $\theta_i$
cosines and sines, respectively):
\begin{equation}
\left( \matrix{ \nu_e \cr  \nu_{\mu} \cr \nu_{\tau} } \right)=
\left( \matrix{ U_{e1} & U_{e2} & U_{e3} \cr U_{\mu 1} & U_{\mu 2} & U_{\mu 3} \cr
                U_{\tau 1} & U_{\tau 2} & U_{\tau 3} } \right) 
\left( \matrix{ \nu_1\cr  \nu_2 \cr \nu_3 } \right),
\end{equation}

where
$$
\left( \matrix{ U_{e1} & U_{e2} & U_{e3} \cr U_{\mu 1} & U_{\mu 2} & U_{\mu 3} \cr
                U_{\tau 1} & U_{\tau 2} & U_{\tau 3} } \right) = 
                \left\{ \matrix{ U,\;\;\; \mbox{\rm for 
                                                      oscillations} 
\cr UV,\;\;\; \mbox{\rm for}\; 
                                         (\beta \beta )_{0 \nu} } \right.
$$ 
and 
\begin{eqnarray*}
U&=&\left( \matrix{ c_1c_3 & c_1 s_3 & s_1 e^{-i \delta} \cr
           -c_2s_3-s_1s_2c_3e^{i \delta} &  c_2c_3-s_1s_2s_3e^{i \delta}  & c_1 s_2 \cr
           -s_2s_3-s_1c_2c_3e^{i \delta} &  -s_2c_3-s_1c_2s_3e^{i \delta} & c_1c_2 } \right), \\
&& \\
V&=&\mbox{\rm diag}(1,e^{i \phi_2} , e^{i(\phi_3+\delta)}).
\end{eqnarray*}

All relevant neutrino experiments have been analyzed assuming that  neutrinos oscillate. 
The results are the following:
\begin{itemize}
\item
From the CHOOZ reactor  \cite{cho}, which measures $\bar{\nu}_e$ disappearance:

\begin{equation}
\sin^2{2 \theta_1} < 0.18\;\;\; \mbox{\rm for}\;\;\; \delta m^2 > 2 \cdot 10^{-3}\; \mbox{\rm eV}^2
\end{equation}

\item
The result of atmospheric neutrino anomaly explained by the 
$\nu_{\mu} \rightarrow \nu_{\tau}$
oscillation, gives \cite{tau}:

\begin{eqnarray}
0.90 & \leq & \cos^4{\theta_1}  \sin^2{2\theta_2} \leq 1.0, \nonumber \\
0 & \leq &  \sin{\theta_1} \leq 0.3,
\end{eqnarray}
with the central value of $\delta m^2$:
\begin{equation}
\delta m^2_{atm} \simeq 3.2 \cdot 10^{-3}\; \mbox{\rm eV}^2.
\end{equation}

\item
Solar neutrino experiments interpreted by 
$\nu_e \rightarrow \nu_{\mu},(\nu_{\tau})$ 
transitions yield
\cite{smi} $(A_{sun}= \cos^4{\theta_1} \sin^2{2\theta_3})$:

\begin{equation} 
0.72 \leq A_{sun} \leq 0.95\;\;\; \mbox{\rm and}\;\;\; \delta m^2_{sun} \simeq 4.42 \cdot 10^{-10} 
\; \mbox{\rm eV}^2,
\end{equation}
in the case of vacuum oscillation (VO);

\begin{equation} 
2 \times 10^{-3}  \leq A_{sun} \leq 10^{-2}\;\;\; \mbox{\rm and} \;\;\;
\delta m^2_{sun} \simeq 5 \cdot 10^{-6} 
\; \mbox{\rm eV}^2,
\end{equation}
in the case of small mixing MSW transition (SMA MSW) 
and 
\begin{equation} 
0.65 \leq A_{sun} \leq 1.0\;\;\; \mbox{\rm and}\;\;\; \delta m^2_{sun} \simeq 2 \cdot 10^{-5} 
\; \mbox{\rm eV}^2,
\end{equation}
in the case of large mixing MSW transition (LMA MSW).
\end{itemize}

Direct kinematical measurement of the $\bar{\nu}_e$ mass from the tritium $\beta $ decay
yield:
\begin{equation}
m(\nu_e)= \left[ |U_{e1}|^2m_1+ |U_{e2}|^2m_2+ |U_{e3}|^2m_3 \right] < m_{\beta} 
\end{equation}
where values of $m_{\beta}$ differ slightly between the two collaborations, 
namely:
$$
 m_{\beta}=3.4 \; \mbox{\rm eV}\; \cite{mai}, \;\;\; \mbox{\rm }\;\;\;
m_{\beta}=2.7\; \mbox{\rm  eV}\; \cite{tro}. 
$$

Combining experimental constraints from atmospheric and solar neutrino
oscillations with the tritium $\beta $ decay limit (Eq.(11)) it is
possible to infer interesting limits on the highest mass eingenvalue and 
on mass differences.

For  three neutrino eigenmasses

\begin{equation}
m_1<m_2<m_3,
\end{equation}

it was found \cite{b2}:

\begin{eqnarray}
\sqrt{\delta m^2_{atm}} & \leq & m_3 \leq \sqrt{m_{\beta}^2+\delta m^2_{atm}}, \nonumber \\
|m_i-m_j| & < & \sqrt{\delta m^2_{atm}}, \;\;\; i,j=1,2,3
\end{eqnarray}
For the present experimental values ($\delta m_{atm}^2 \simeq (1.5 \div 6.0)
\cdot 10^{-3}\; \mbox{\rm eV}^2$ (90 \% C.L.) \cite{tau} we have:

\begin{eqnarray}
0.04\; \mbox{\rm eV}& < & m_3 < 2.7\; \mbox{\rm eV}, \\
|m_i-m_j|& <& 0.08\;  \mbox{\rm eV},\;\;\; i,j=1,2,3.
\end{eqnarray}

These  constraints are  satisfied by two possible mass spectra given in Fig.1.
In addition the total scale  for neutrino masses is not fixed.

\begin{figure}
\epsfig{figure=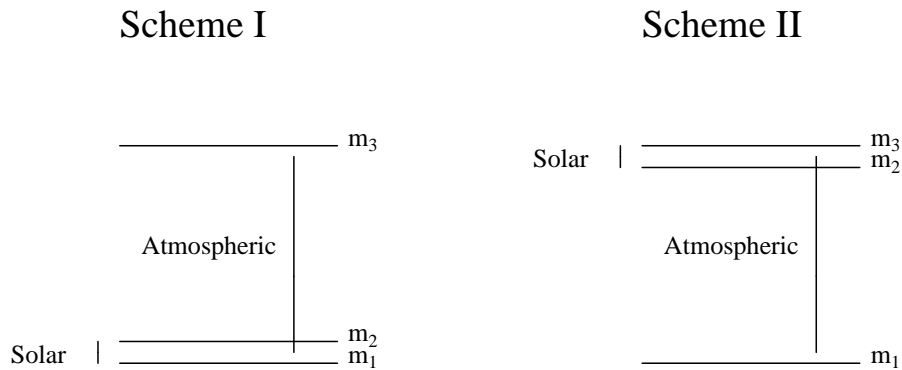, height=2 in}
\caption{Two possible neutrino mass spectra which can describe the oscillation data.
Scheme I describes 
the atmospheric anomaly with  $\delta m^2_{31} \simeq \delta m^2_{32} = \delta m^2_{atm}$
and  solar
neutrino oscillations with $\delta m^2_{21}=\delta m^2_{sun}$. 
Scheme II describes 
the atmospheric anomaly with  $\delta m^2_{31} \simeq \delta m^2_{21} = \delta m^2_{atm}$
and  solar
neutrino oscillations with $\delta m^2_{32}=\delta m^2_{sun}$.}
\end{figure}
The lowest mass m$_1$ can be close to zero or 
much higher, up to 2.7 eV. 
\begin{enumerate}
\item 
In the first case the mass spectrum can be  hierarchical with  
$m_3>> m_2 > m_1$ (Fig.1;
scheme I) or $m_3 > m_2 >> m_1$ (Fig1; scheme II). 
\item
In the second situation, masses are
almost degenerate $m_1 \simeq m_2 \simeq m_3$, but schemes I and II, 
which describe solar and atmospheric 
anomalies, are the same. Since cosmological estimations give \cite{gel}

\begin{equation}
\sum\limits_{\nu} m_{\nu} \leq 5\;  \mbox{\rm eV},
\end{equation}

all three neutrinos have  masses close to (or below)  1.7 eV.
\end{enumerate}
 
\section{Constraints on the nature of neutrinos}

Oscillations of Dirac and Majorana neutrinos share the same description.
Bounds on mixing angles Eqs.(5-10) do not feel their 
character, contrary to the effective mass (Eq.(2,3)). For a Dirac neutrino 
$<m_{\nu}>=0$. For a Majorana neutrino  \cite{doi}:

\begin{equation}
|<m_{\nu}>|= \left| |U_{e1}|^2m_1+ |U_{e2}|^2m_2 e^{2i \phi_2} + |U_{e3}|^2m_3e^{2i \phi_3}
\right|.
\end{equation} 

The U matrix is unitary, so only two elements of $\left| U_{ei}\right|$ are
independent.

Let us bound $ | U_{e2}|^2$ and $ | U_{e3}|^2$.
From Eqs.(5) and (6) we can find $(x=\cos^2{\theta_1})$
\begin{equation}
0 \leq | U_{e3}|^2 \equiv (1-x) \leq 0.05.
\end{equation}

The constraints on  $|U_{e2}|^2$ are obtained from Eqs.(8),(9) and (10). As the $A_{sun}$
amplitude depends on $\sin^2{2 \theta_3}$, $\sin^2{\theta_3}$ and $\cos^2{\theta_3}$
are not fixed independently. Two values for $\theta _3$ 
in the range $0< \theta_3 < \frac{\pi}{4}$
and $\frac{\pi}{4} < \theta_3 <\frac{\pi}{2}$ can be inferred from bounds on 
$\sin^2{2 \theta_3}$, giving two possible values
of $|U_{e2}|^2$. We will call them small and large  $\sin^2{\theta_3}$.

The  vacuum oscillation Eq.(8) yields:
\begin{enumerate}
\item
For small $\sin^2{\theta_3}$:

$$
\frac{x}{2} \left( 1- \sqrt{1-\frac{0.72}{x^2}} \right) \leq
{|U_{e2}|}_S^{2(VO)} \leq \left\{ \matrix{ \frac{x}{2} &  0.9 \leq x^2 
\leq 0.95  \cr
\frac{x}{2} \left( 1-\sqrt{1-\frac{0.95}{x^2}} \right) &  0.95 \leq x^2 
\leq 1.0  } \right. 
$$


\item
For large $\sin^2{\theta_3}$:

$$
\frac{x}{2} \left( 1+ \sqrt{1-\frac{0.72}{x^2}} \right) \geq
|U_{e2}|_L^{2(VO)} \geq \left\{ \matrix{\frac{x}{2} &  0.9 \leq x^2 \leq 0.95  \cr
\frac{x}{2} \left( 1+\sqrt{1-\frac{0.95}{x^2}} \right) &  0.95 \leq x^2 \leq 1.0  } 
\right. 
$$

\begin{equation}
\end{equation}
\end{enumerate}
Therefore, $0.95 \leq x \leq 1.0$ leads to: 
\begin{eqnarray}
0.24 & \leq & {|U_{e2}|}_S^{2(VO)} \leq 0.48, \\
0.48 & \leq & {|U_{e2}|}_L^{2(VO)} \leq 0.76.
\end{eqnarray}

In a similar way, bounds from the  SMA and LMA solutions to the solar anomaly
are:
\begin{enumerate}
\item
\begin{eqnarray}
0.0005 & \leq & {|U_{e2}|}_S^{2(SMA)} \leq 0.0026, \\
0.947 & \leq & {|U_{e2}|}_L^{2(SMA)} \leq 0.999.
\end{eqnarray}
for SMA MSW, and:
\item
\begin{eqnarray}
0.204 & \leq & {|U_{e2}|}_S^{2(LMA)} \leq 0.48, \\
0.48 & \leq & {|U_{e2}|}_L^{2(LMA)} \leq 0.8,
\end{eqnarray}
for LMA MSW.
\end{enumerate}

Equipped with bounds on $|U_{e2}|^2$ and $|U_{e3}|^2$, we are able to explore the
meaning of Eqs.(2,3).
 
\subsection{Hierarchical neutrino mass scenario}

First, let us consider the mass scheme I from Fig.1.
Since  $m_1 \simeq 0$:

\begin{eqnarray}
m_2 & = & \sqrt{\delta m^2_{sun}}=\left\{ \matrix{ 2.1 \cdot 10^{-5}\; eV & (VO) \cr
                                                 2.2 \cdot 10^{-3}\;  eV & (SMA) \cr
                                                 4.5 \cdot 10^{-3}\;  eV & (LMA) , } 
\right. \\
m_3&=&\sqrt{\delta m^2_{atm}}=0.057\; eV
\end{eqnarray}

Using the value of $m_3$, the experimental data on  ($\beta \beta $)$_{0 \nu}$
translate into the following:

$$
\frac{|<m_{\nu}>|}{m_3} =\left| |U_{e2}|^2\frac{m_2}{m_3}+ |U_{e3}|^2 e^{2i (\phi_2-\phi_3)}
\right| \leq
\left\{ \matrix{3.5 & \mbox{\rm (present bound)}  \cr
0.18  & \mbox{\rm (GENIUS)} } \right. 
$$
\begin{equation}
\end{equation}

On the other hand, 
even for the large $\sin^2{\theta_3}$ solution, where $U_{e2} \simeq 1$,
we get:
\begin{equation}
\left( \frac{|<m_{\nu}>|}{m_3} \right)_{max}=\frac{m_2}{m_3}+ |U_{e3}|^2 \leq 0.13.
\end{equation}

This means that even the future sensitivity of GENIUS will not be enough to discern
neutrino nature.

Let us explore now  the hierarchical mass scheme II from Fig.1:
\begin{eqnarray}
m_2 &=& \sqrt{\delta m^2_{atm}} =0.057\; \mbox{\rm eV}, \nonumber \\
m_3 &=& \sqrt{\delta m^2_{atm}+\delta m^2_{sun}} \simeq 0.057\; \mbox{\rm eV.}
\end{eqnarray}
This again can be transformed into:
\begin{equation}
\frac{|<m_{\nu}>|}{m_3} =\left| |U_{e2}|^2+ |U_{e3}|^2 e^{2i (\phi_2-\phi_3)} \right|
\leq
\left\{ \matrix{3.5 & \mbox{\rm (present bound)}  \cr
0.18  & \mbox{\rm (GENIUS)} } \right. 
\end{equation}

From unitarity of the U  matrix follows that the present bound is always
satisfied. The result of  GENIUS  is much more interesting. We can find
the minimal values of the effective mass

\begin{equation}
\left( \frac{|<m_{\nu}>|}{m_3} \right)_{min} = \left| |U_{e2}|^2-
|U_{e3}|^2 \right| ,
\end{equation}

namely:

\begin{equation}
{\left( \frac{|<m_{\nu}>|}{m_3} \right)}_{min} =
\left\{ \matrix{ 0.21 & \mbox{\rm VO, small}\; \sin^2{\theta_3}, \cr
                 0.42 & \mbox{\rm VO, large}  \sin^2{\theta_3}, \cr
                < 0.05 & \mbox{\rm SMA, small}  \sin^2{\theta_3}, \cr
                 0.9 & \mbox{\rm SMA, large}  \sin^2{\theta_3}, \cr
                 0.18 & \mbox{\rm LMA, small}  \sin^2{\theta_3}, \cr
                 0.43 & \mbox{\rm LMA, large}  \sin^2{\theta_3} } . \right.
\end{equation}
For VO
and LMA MSW solutions of the solar neutrino problem, a negative result of the GENIUS 
experiment would mean that neutrinos must be Dirac particles,
since our bounds exceed those infered by this collaboration. 
For the SMA MSW solution, conclusions depend on the values of $ \sin^2{\theta_3}$. 
Next experiments 
(SNO, BOREXINO) which will be able to discern types of oscillations can give informations on
$ \sin^2{\theta_3}$.


\subsection{Degenerate neutrino mass scenario}

We assume that neutrino
mass is much larger than $\sqrt{\delta m^2_{atm}}$. 
Let it be   that $\sum\limits_{\nu} m_{\nu}
\simeq 5$ eV, so $m_{\nu} \simeq 1.7$ eV. 
Now, the ($%
\beta \beta $)$_{0 \nu}$ bound has the form:

$$
\frac{|<m_{\nu}>|}{m_{\nu}} =\left| |U_{e1}|^2+|U_{e2}|^2 e^{2i \phi_2} + |U_{e3}|^2 
e^{2i \phi_3} \right| \leq
\left\{ \matrix{0.12 & \mbox{\rm (present bound)}  \cr
0.006  &  \mbox{\rm  (GENIUS)} } \right. 
$$
\begin{equation}
\end{equation}

The discussion is more difficult, since we do not have any information about the CP
violating phases $\phi_1,\phi_2$. So, let us assume that CP is conserved and consider
all possible $CP=\pm i$ $(e^{i \phi_i}=\pm 1)$ phases for neutrinos. 
As only relative CP phases have
physical consequences, we take $\eta_{CP}(\nu_1)=+i$. There are four different
arrangements: (A)  $\eta_{CP}(\nu_1)=\eta_{CP}(\nu_2)=\eta_{CP}(\nu_3)=+i$, 
(B)  $\eta_{CP}(\nu_1)=-\eta_{CP}(\nu_2)=-\eta_{CP}(\nu_3)$,
(C)  $\eta_{CP}(\nu_1)=\eta_{CP}(\nu_2)=-\eta_{CP}(\nu_3)$
and (D) 
 $\eta_{CP}(\nu_1)=-\eta_{CP}(\nu_2)=\eta_{CP}(\nu_3)$.
In the case,  CP is broken, the condition (34) looks more complicated, but qualitatively
our discussion will not change. The four CP conserving cases give extreme
values for $|<m_{\nu}>|$, namely:
 
\begin{eqnarray}
(A): \hspace{5 mm} \frac{<m_{\nu}>}{m_{\nu}} & = & 1, \\
(B): \hspace{5 mm} \frac{<m_{\nu}>}{m_{\nu}} & = & \left| 1-2|U_{e2}|^2-
2|U_{e3}|^2 \right|, \\
(C): \hspace{5 mm} \frac{<m_{\nu}>}{m_{\nu}} & = & \left| 1-2|U_{e3}|^2 \right|, \\
(D): \hspace{5 mm} \frac{<m_{\nu}>}{m_{\nu}} & = & \left| 1-2|U_{e2}|^2 \right|, 
\end{eqnarray}
Taking all possible values of the mixing matrix elements (Eqs. (17)-(25))
several interesting conclusions come to mind.

1) If neutrinos are Majorana particles and CP is conserved, already the present bound
(Eq.(34)) is not  satisfied in the cases (A) and (C) for $m_{\nu} =1.7$ eV,
independently of
the oscillation mechanism for the solar neutrino deficit. The lower bound
on  the masses for which this statement is true is approximately 0.22 eV. 
This means that
almost degenerate neutrinos cannot have the same CP parities \cite{raj}, and the CP
parity of the heaviest neutrino cannot be opposite to the CP eigenvalues of
the lighter ones.

Let us discuss cases (B) and (D).

2) For the vacuum oscillation mechanism, the only interesting possibility
is the (D) case, where $\left( \frac{|<m_{\nu}>|}{m_{\nu}} \right)_{min}  \simeq 0.040$ 
for small $\sin^2{\theta_3}$. Other oscillation scenarios, namely with large 
$\sin^2{\theta_3}$ and the case (B) allow  $\left( \frac{|<m_{\nu}>|}{m_{\nu}} \right)_{min} 
\simeq 0$.

3) For SMA MSW, 
both cases (B) and (D) give a minimal value of $\frac{<m_{\nu}>}{m_{\nu}}$  
greater than 0.9. 
The present bound (Eq.(34)) is not satisfied, neutrinos would have to be
Dirac particles.
Or,  by opposite, if neutrinos are almost degenerate
Majorana particles with $m_{\nu}>0.22$ eV, the SMA MSW mechanism is not a proper
explanation of the solar neutrino deficit \cite{geo}.

4) The LMA MSW case is similar to the VO case, because the ranges  of the U matrix 
elements are comparable (Eqs.(20)-(21) and  Eqs.(24-25)).

\section{Conclusions}

We have combined data on solar and atmospheric neutrino oscillations with the present
and future bounds on ($\beta \beta $)$_{0 \nu}$.

Present ($\beta \beta $)$_{0 \nu}$ experiments imply:

\begin{itemize}
\item None of the hierarchical schemes I and II is able to help us to determine the 
nature of neutrinos. 
\item A degenerate scheme with Majorana neutrinos of (A) or (C) parities is excluded
if $m_{\nu}>0.22$ eV.
\item Degenerate  Majorana neutrinos with $m_{\nu} > 0.22$ eV and both (B),(D) parity
assignments  combined with the SMA MSW are excluded.
\end{itemize}

The future ($\beta \beta $)$_{0 \nu}$ experiment GENIUS gives a chance 
to see  whether:
\begin{itemize}
\item  Majorana neutrinos of the  hierarchical scheme II fit the data, which is not the 
case of scheme I;
\item  the VO  mechanism  with degenerate neutrinos, small  $\sin^2{\theta_3}$ 
and (D) parity assignments is correct.
\end{itemize}

\begin{ack}
This work was supported by Polish Committee for Scientific Research under 
Grants Nos.  2P03B08414  and 2P03B04215. 
J.G. would like to thank the Alexander von Humboldt-Stiftung for a fellowship.
\end{ack}


\begin{thebibliography}{99}
\bibitem{zr} M. Zra\l ek, Acta Phys. Pol. {\bf 28} (1997) 2225.
\bibitem{li} L.F. Li, F. Wilczek, \pr{D25} (1982) 143;
B. Kayser, R.E. Schrock, \pl{\bf B112} (1982) 137;y
B. Kayser, \pr{D26} (1982) 1662.
\bibitem{bb} L. Baudis et. al., hep-ex/9902014.
\bibitem{kla} H.V. Klapdor-Kleingrothaus, J. Hellmig, M. Hirsch, J. Phys.
{\bf G24} (1998) 484; J. Hellmig, H.V. Klapdor-Kleingrothaus, \zp{A359}(1997)351;
 H.V. Klapdor-Kleingrothaus, M. Hirsch, \zp{A359}(1997)361;
C.E. Aalseth et. al., Nucl. Phys. (Proc. Suppl.) {\bf 70} (1999) 236;
X. Sarazin (NEMO Coll.) ibid. {\bf 70} (1999) 239; V.D. Ashitkov et. al., ibid. {\bf 70} (1999) 233;
E. Fiorini, Phys. Rep. {\bf 307} (1998) 309;
H.V. Klapdor-Kleingrothaus, hep-ex/9907040.
\bibitem{previous} S.T. Petcov, A. Yu. Smirnov, Phys. Lett. {\bf B322} (1994) 109;
S.M. Bilenky, A. Bottino, C. Giunti, C.W. Kim, Phys. Lett. {\bf B356} (1995) 273;
S.M. Bilenky, C. Giunti, C.W. Kim, M. Monteno, Phys. Rev. {\bf D57} (1998) 6981;
S.M. Bilenky, C. Giunti, W. Grimus, hep-ph/9809368;
F. Vissani, hep-ph/9708483; hep-ph/9904349;
hep-ph/9906525;
\bibitem{b1} V. Barger and K. Whisnant, hep-ph/9904281.
\bibitem{raj} R. Adhikair, G. Rajasekaran, hep-ph/9812361.
\bibitem{geo} H. Georgi, S.L. Glashow, hep-ph/9808292;
H. Minakata, O. Yasuda, \pr{D56} (1997) 1692.  
\bibitem{ten} S.M. Bilenky, C. Giunti, hep-ph/9904328; C. Giunti,
hep-ph/9906275.
S.M. Bilenky, C. Giunti, W. Grimus, B. Kayser, S.T. Petcov, hep-ph/9907234.
\bibitem{eleven} G.C. Branco, M.N. Rebelo, J.I. Silva-Marcos, Phys. Rev. Lett. {\bf 82} (1999) 682.
\bibitem{review} see review papers:

S.M. Bilenky, C. Giunti, W. Grimus, hep-ph/9812360;
Europhysics Neutrino Oscillation Workshop 98, hep-ph/9906251;
P. Fisher, B. Kayser, K.S. McFarland, hep-ph/9906244;
J. Ellis, hep-ph/9907458.
\bibitem{lsnd} C. Athanassopoulos et al., \prl{77}(1996)3082;
\pr{C54}(1996)268; \pr{C58}(1998)2489; \prl{81}(1998)1774; S.J. Yellin,
hep-ex/9902012.
\bibitem{con} B. Zeitnitz et al., Prog. Part. Nucl. Phys., {\bf 40} (1998)
169; B. Armbruster et. al., \pr{C57} (1998) 3414; E. Eitel and B. Zeitnitz,
hep-ex/9809007;
for MiniBooNE see http://www.neutrino.lanl.gov/BooNE/.
\bibitem{aq} F. del Aquila, M. Zra\l ek, \np{B447}(1995)211;
Acta Phys. Pol. {\bf B27} (1996) 971.
\bibitem{cho} CHOOZ Collaboration, M. Apollonio et. al., \pl{B420}(1998)397.
\bibitem{tau} K. Scholberg, hep-ex/9905016.
\bibitem{smi} M.B. Smy, hep-ex/9903034;
J.N. Bahcall, P.I. Krastev, A. Yu. Smirnow, \pr{D58}(1998)096016;
hep-ph/9905220.
\bibitem{mai} C. Weinheimer (Mainz), Talk at Ringberg Euroconference ``New Trends in Neutrino
Physics'', May 1998;
H. Berth et al., Prog. Part. Nucl. Phys. {\bf 40} (1998) 353;
C. Weinheimer et. al., \pl{\bf B300} (1993) 210.
\bibitem{tro} V.M. Lobashev (Troitsk), Talk at Ringberg Euroconference ``New Trends in Neutrino
Physics'', May 1998; V.M. Lobashev, Prog. Part. Nucl. Phys. {\bf 40} (1998) 337;
A.I. Belesev et.al., \pl{\bf B350} (1995) 263.
\bibitem{b2} V. Barger, T.J. Weiler, K. Whisnant, \pl{B442}(1998)255.
\bibitem{gel} see e.g. G.G. Raffelt, hep-ph/990271;
G. Gelmini, hep-ph/9904369.
\bibitem{doi} see e.g. M. Doi, T. Kotani, E. Takasugi, Prog. Theor. Phys.
(supplement) {\bf 83} (1985) 1.
\end{thebibliography}
\end{document}